\documentclass{czjphys}         
\RequirePackage{epsfig}

\begin{document}
\title{Non-Hermitian Quantum Theory and its Holomorphic Representation: Introduction and Applications}  
%
\authori{Frieder Kleefeld\,\footnote{E-mail: {\sf kleefeld@cfif.ist.utl.pt}\,, URL: {\sf http:/$\!$/cfif.ist.utl.pt/$\sim$kleefeld/}}}      \addressi{Centro de F\'{\i}sica das Interac\c{c}\~{o}es Fundamentais (CFIF), Instituto Superior T\'{e}cnico,\\ Av.\ Rovisco Pais, 1049-001 Lisboa, Portugal}
\authorii{}     \addressii{}
\authoriii{}    \addressiii{}
\authoriv{}     \addressiv{}
\authorv{}      \addressv{}
\authorvi{}     \addressvi{}
%
\headauthor{Frieder Kleefeld}            
\headtitle{Non-Hermitian Quantum Theory and its Holomorphic Representation \ldots}             
\lastevenhead{Frieder Kleefeld: Non-Hermitian Quantum Theory and its Holomorphic Representation \ldots} 
\refnum{A}
\daterec{XXX}    
\issuenumber{?}  \year{2004}
\setcounter{page}{1}
\maketitle

\begin{abstract} This article contains a short summary of an oral presentation in the 2nd International Workshop on ``Pseudo-Hermitian Hamiltonians in Quantum Physics'' (14.-16.6.2004, Villa Lanna, Prague, Czech Republic). The purpose of the presentation has been to introduce a non-Hermitian generalization of pseudo-Hermitian Quantum Theory (QT) allowing to reconcile the orthogonal concepts of causality, Poincar\'{e} invariance, analyticity, and locality. We conclude by considering interesting applications like non-Hermitian supersymmetry.
\end{abstract}

\section{Introduction}
This short overview wants to outline just some main issues which made part of our comprehensive oral presentation on June 15, 2004. A detailed survey of the presented material can be found in Ref.\ \cite{Kleefeld:2004jb} containing a lot of relevant references. During the presentation we tried to propose a formalism called Non-Hermitian Quantum Theory (NHQT) or (Anti)Causal Quantum Theory (A)CQT (See e.g.\ Refs.\ \cite{Kleefeld:2004jb,Kleefeld:2003dx,Kleefeld:2002au,Kleefeld:2001xd,Kleefeld:1998yj}!) allowing to quantize consistently systems decribed by non-Hermitian Hamilton operators and to reconcile the seemingly orthogonal concepts of causality, Poincar\'{e} invariance, analyticity, locality. Also we identified the holomorphic representation of complex analysis as the spacial representation of NHQT.
\section{Hermitian and Pseudo-Hermitian Quantum Theory}
One axiom of Quantum Mechanis (QM), as we find it presently in text books, states that  observables are represented by {\em Hermitian} operators $A$, with functions of observables being represented by the corresponding functions of the operators. The Hermiticity imposed hereby on operators $A$ respresenting observables is guided by the belief that expectation values  $\left<A\right>$ in physically acceptable theories admitting a probability interpretation should be {\em real-valued}, while probabilities are understood to be {\em real-valued positive} numbers in the interval from zero to one. Historically it appeared as a surprise that in the so-called (normal) ``Lee-model'' proposed 1954 \cite{Lee:1954iq} in a certain range of the coupling parameter space the Hermitian renormalized Hamilton operator was corresponding to an explicitly non-Hermitian non-renormalized Hamilton operator, which seemed to be conflicting with the above mentioned axioms of QM. Puzzled by his observation T.D.\ Lee suggested \cite{Lee:1954iq} that the non-Hermitian non-renormalized Hamilton operator and the Hermitian renormalized Hamilton operator might be related by a non-unitary similarity transform. In June, 1955, W.~Pauli~\cite{Pauli:1956} in collaboration with G.\ K\"all\'{e}n \cite{kaellen1955} came to the conclusion \cite{Pauli:1956} that the Lee-model --- in the particular range of the coupling parameter space --- {\em ``$\ldots$ leads to a contradiction with the concept of physical probability (indefinite metric of the Hilbert space) $\ldots$ connected with the appearance of new discrete stationary states whose contribution to the conserved sum of `probabilities' is negative (`ghosts') $\ldots$''}. Discovering a metric operator $\eta$~\cite{Pauli:1943} of {\em indefinite metric} \cite{pandit1959,nagy1960b,Kleefeld:Nakanishi:1972pt} in the Lee-model they realized that the Hamilton operator $H$ of the Lee-model is actually {\em pseudo-Hermitian} \cite{heis1961,pauli1958,pandit1959} ($H=\eta^{-1} H^+ \eta$) admitting negative probabilities as already considered by P.A.M.\ Dirac in 1941 \cite{Dirac:1942} (See also \cite{Pauli:1943,mueckenheim1}!) and also biorthogonal eigenstates (``complex ghosts'' \cite{Kleefeld:Nakanishi:1972pt}) of zero traditional norm with complex-valued pairs of mutually complex-conjugate energy eigenvalues. After W.\ Heisenberg's conjecture \cite{duerr1969} {\em ``$\ldots$ that the existence of a unitary S-matrix for physical states will be sufficient to guarantee the usual quantum-mechanical probability interpretation $\ldots$''}, W.~Pauli~\cite{pauli1958} and W.~Heisenberg \cite{heis1957} (See also Ref.\ \cite{ascoli1958}!) were the first to understand the conditions under which the S-matrix for {\em pseudo-unitary} \cite{pauli1958,pandit1959} systems will admit a probability interpretation even for QTs containing simultaneously (complex) ghosts and eigenstates with strictly real energy eigenvalues. Unfortunately --- as we will argue below --- a QT allowing the interaction of states with real and complex energy {\em violates causality, Poincar\'{e} invariance, analyticity and locality}. We will illustrate, how these deficiencies of Hermitian QT can be fortunately removed in (A)CQT, yet at the price of introducing a {\em complex probability concept} allowing $\left<\!\left<\psi\right|\right.\!\left.\psi\right>_\eta \equiv \left<\psi\right| \! \left. \eta \,\, \psi\right>$ to be complex, and hence also $\eta \not=\eta^+$, leading necessarily {\em beyond} the suggestion made in Section 3 of Ref.~\cite{znojil2004pt}~\footnote{Since $\simeq$ 1980 there has received the class of pseudo-Hermitian (mostly non-Hermitian) Hamilton operators with PT-symmetry particular attention due to their still surprising, yet nearly forgotten feature of admitting some real spectrum and a probability interpretation (See e.g.\ \cite{pseudo2003,Bender:1998ke,Znojil:2001,Znojil:2004xw}!). 
A review provided in 2001 by M.\ Znojil as a preprint {\sf math-ph/0104012} (unfortunately published very delayed in 2004 \cite{Znojil:2001}) containing a lot of important references to related work showed that the new research field had again, yet independently reached a similar level of understanding as the disciples of the Lee-model around 1970. In Ref.\ \cite{Znojil:2001} one does not only find discussed the concept of {\em pseudo-Hermiticity}, {\em pseudo-unitarity} and {\em indefinite metric} (i.e.\ $P$). It is also explained under the headline {\em Spontaneous Broken {\cal PT} Symmetry} the situation, when the Hamilton operator develops complex ghosts. Some important new contributions to the field are mentioned in Ref.\ \cite{Kleefeld:2004jb}.}.
\section{(Anti)Causal Quantum Theory}
\subsection{The (Anti)Causal Harmonic Oscillator} \label{acho1}
The simplest relativistic field-theoretic extension of {\em pseudo-Hermitian} QT is a non-interacting Quantum Theory of a causal Klein-Gordon (KG) field $\phi(x)$ with complex mass $M := m - \frac{i}{2} \, \Gamma$ and the anti-causal KG field $\phi^+(x)$ with complex-conjugate mass $M^\ast$, or the respective non-interacting Quantum Theory of a causal Dirac field $\psi(x)$ with complex mass $M$ and the anti-causal Dirac field $\psi^c(x)$ with complex mass $\bar{M}=\gamma_0 M^+ \gamma_0$ \footnote{The (Anti)Causal KG theory, the Lagrangean of which was for the first time denoted 1959 by M.~Froissart \cite{froi1959}, was studied in 1970 by T.D.\ Lee \& G.C.\ Wick \cite{Lee:iw} in the context of vector fields and received a more thorough investigation by N.\ Nakanishi \cite{Kleefeld:Nakanishi:wx,Kleefeld:Nakanishi:1972pt} under the name ``Complex-Ghost Relativistic Field Theory'', while the formalism$\;\!$was$\;\!$1999-2000 independently rederived by the author (See e.g.\ Refs.\ \cite{Kleefeld:2002au,Kleefeld:2001xd}). The (Anti)Causal Dirac theory was for the first time considered in 1970 by Lee \& Wick \cite{Lee:iw} and later independently rederived by the author (See e.g.\ Refs.\ \cite{Kleefeld:2002au,Kleefeld:1998yj}!).}. The Lagrangeans for {\em neutral} (anti)causal KG and Dirac fields, which lead to the causal and anticausal KG equations $(\,\partial^2 + M^2) \,\phi (x) = 0$ and $(\,\partial^2 + M^{\ast \, 2}) \,\phi^+ (x) = 0$, and to the (anti)causal Dirac equations  $( i \!\not\!\partial - \, M ) \; \psi (x) = 0$, $( i \!\not\!\partial - \, \bar{M} ) \; \psi^c (x) = 0$, $\overline{\psi^c} (x) \; ( - \, i \! \stackrel{\;\,\leftarrow}{\partial\!\!\!/} - \, M ) = 0$ and $\bar{\psi} (x) \; ( - \, i \! \stackrel{\;\,\leftarrow}{\partial\!\!\!/} - \, \bar{M} ) = 0$, are \cite{Kleefeld:2004jb}:
\begin{eqnarray}
{\cal L} (x) & = & \frac{1}{2}\, \Big( (\partial \,\phi (x) )^2  - \, M^2 \, (\phi (x))^2 \, \Big)\; + \;
\frac{1}{2}\, \Big( (\partial \,\phi^+ (x) )^2  - \, M^{\ast \, 2} \,
(\phi^+ (x))^2 \Big) \; , \nonumber \\
{\cal L} (x) & = & \frac{1}{2} \,\Big( \; \overline{\psi^c} (x) \, ( \frac{1}{2} \, i \! \stackrel{\;\,\leftrightarrow}{\partial\!\!\!/} \! -  M ) \, \psi (x) \; + \; \bar{\psi} (x) \, ( \frac{1}{2} \, i \! \stackrel{\;\,\leftrightarrow}{\partial\!\!\!/} \! -  \bar{M} ) \, \psi^c (x) \; \Big) \; . \label{kglag1}
\end{eqnarray}
We want to mention without going into the details addressed e.g.\ in Ref.\ \cite{Kleefeld:2004jb} that the formulation of complex mass Dirac theory requires a generalized spinor concept relying on generalized Lorentz-boosts for (Fermionic and Bosonic) systems with complex mass.
As discussed in Ref.\ \cite{Kleefeld:2004jb} standard {\em real-equal-time} canonical quantization leads after Legendre transform to the following KG/Dirac ($\pm$) {\em diagonal} Hamilton operator $H =
\int\! \!
\frac{d^3 p}{(2\pi )^3 \, 2\, \omega \, (\vec{p}\,)}
\,
\frac{\omega \,(\vec{p}\,)}{2} [ c^+ (\vec{p}\,) , a (\vec{p}\,)]_\pm +
 \int\! \! \frac{d^3 p}{(2\pi )^3 \, 2\, \omega^\ast (\vec{p}\,)}
\, \frac{\omega^\ast (\vec{p}\,)}{2} [a^+ (\vec{p}\,) , c\, (\vec{p}\,)]_\pm $ in 
combination with the {\em non-vanishing} (anti)commutation relations ${[ \, a \, (\vec{p}\,) , c^+ (\vec{p}^{\,\,\prime}) \, ]_\mp} = (2\pi)^3 \, 2 \; \omega \,(\vec{p}\,)\, \delta^{\, 3} (\vec{p} - \vec{p}^{\,\,\prime}\,)$ and ${ [ \, c \, (\vec{p}\,) , a^+ (\vec{p}^{\,\,\prime}) \, ]_\mp } = (2\pi)^3 \, 2 \; \omega^\ast(\vec{p}\,)\, \delta^{\, 3} (\vec{p} - \vec{p}^{\,\,\prime}\,)$. The system is studied much easier as a 1-dimensional quantum-mechanical (anti)causal Harmonic Oscillator described by the Hamilton Operator \cite{Kleefeld:2004jb}
$H = H_C + H_A = \frac{\omega}{2}\, [c^+, a\, ]_\pm + \frac{\omega^\ast}{2} \, [a^+, c\, ]_\pm$ ($\pm$ for Bosons/Fermions \footnote{The Fermionic case we tend to denote by $H = \frac{1}{2} \omega [d^+, b ] + \frac{1}{2} \omega^\ast [b^+, d ]$ with $\{\,b, d^+ \} = 1$ etc.\,.}). For this {\em pseudo-Hermitian} Harmonic Oscillator the underlying causal and anticausal Hamilton operators $H_C$ and $H_A$  are related by $H_A = H^+_C$, yielding the Hermiticity relation $H=H^+$. That this relation is actually {\em not reflecting Hermiticity}, but {\em pseudo-Hermiticity} can be seen from the underlying (anti)commuation relations, which contain an {\em indefinite metric}:
\begin{eqnarray} \left( \begin{array}{cc} {[c,c^+]_\mp} & {[c,a^+]_\mp} \\
{[a,c^+]_\mp} & {[a,a^+]_\mp} \end{array}\right) & = & \left( \begin{array}{cc} 0 & 1 \\
1 & 0 \end{array}\right) \;\; = \;\; \mbox{``indefinite metric''} \; , \nonumber \\[1mm]
\left( \begin{array}{cc} {[c,c]_\mp} & {[c,a]_\mp} \\
{[a,c]_\mp} & {[a,a]_\mp} \end{array}\right) \;\; & = & \left( \begin{array}{cc} {[c^+,c^+]_\mp} & {[c^+,a^+]_\mp} \\
{[a^+,c^+]_\mp} & {[a^+,a^+]_\mp} \end{array}\right) \;\; = \;\; \left( \begin{array}{cc} 0 & 0 \\
0 & 0 \end{array}\right) \; .
\end{eqnarray}
That the causal time-development of a system in a space-time characterized by the indefinite Minkowski-metric ($+$,$-$,$-$,$-$) is described by (anti)commutation relations containing also an indefinite metric should not surprise!
Obviously there holds also $[H_C,H_A]=0$. The Hamilton operator is diagonalized by the (normalized) {\em normal} right eigenstates $\left|n,m\right>= (c^+)^n (a^+)^m \left|0\right>/\sqrt{n!\,m!}$ and left eigenstates $\left<\!\left<\,n,m\right|\right.=\left<\!\left<\,0\right|\right.\! c^m \, a^n /\sqrt{m!\,n!}$ (Bosons: $n,m\in{\rm I\!N}$${}_0$\,; Fermions: $n,m\in\{0,1\}$)
being solutions of the equations $(H-E_{\,n,m}) \left|n,m\right> = 0$ and $\left<\!\left< n,m\right|\right.\! (H-E_{\,n,m}) = 0$
for the eigenvalues $E_{n,m} = \omega  (n \pm \frac{1}{2}) + \omega^\ast (m \pm \frac{1}{2})$. The eigenvalues $E_{n,n}$ are obviously {\em real}, while the eigenvalues $E_{m,n}$ and $E_{n,m}$ form a {\em complex conjugate} pair of ``complex ghosts'' for $n\not=m$ which arises typically for the case of broken ``PT''-symmetry.
The {\em (bi)orthogonal} eigenstates are complete, i.e. $\left<\!\left< n^\prime,m^\prime\right|\right.\!\left.n,m\right> = \delta_{n^\prime n} \, \delta_{m^\prime m}$, $\sum_{n,m} \left|n,m\right>\left<\!\left<n,m\right|\right. = \mbox{\bf 1}$.
\subsection{(Anti)Causal Schr\"odinger Theory and the Complex Probability Concept}
According to Ref.\ \cite{Kleefeld:2004jb} there are as much as eight time-dependent Schr\"odinger-like equations describing the time evolution of states in (A)CQT, i.e.\ four causal equations $\big(i\,\partial_t - H\big) \big| \psi (t) \big> = 0$, $\big(i\,\partial_t + H\big)\big| \tilde{\psi} (t) \big>  = 0$, $\big<\!\big< \tilde{\psi} (t) \big|\big(- i\stackrel{\leftarrow}{\partial}_t - \stackrel{\,\leftarrow}{H} \big) = 0$, $\big<\!\big< \psi (t) \big| \big( - i\stackrel{\leftarrow}{\partial}_t + \stackrel{\,\leftarrow}{H}\big) = 0$, and four anticausal equations $\big< \psi (t) \big| \big(-\,i\stackrel{\leftarrow}{\partial}_t - \, H^+\big) = 0$, $\big< \tilde{\psi} (t) \big| \big(-\,i\stackrel{\leftarrow}{\partial}_t + \, H^+\big) = 0$, $\big(i\,\partial_t - ( \stackrel{\,\leftarrow}{H})^+\big) \big| \tilde{\psi} (t) \big>\!\big> = 0$, $\big(i\,\partial_t + ( \stackrel{\,\leftarrow}{H} )^+\big) \big| \psi (t) \big>\!\big> = 0$.
The four causal equations are related to corresponding four anticausal equations by Hermitian conjugation.
From the four causal equations we can derive the continuity equations $\partial_t \big<\!\big< \tilde{\psi} (t) \big| \psi (t) \big> = - \, i \, \big<\!\big< \tilde{\psi} (t) \big|\big(H \,- \stackrel{\,\leftarrow}{H}\! \big)\big| \psi (t) \big>$ and
$\partial_t \big<\!\big< \psi (t) \big| \tilde{\psi} (t) \big> = + \, i \, \big<\!\big< \psi (t) \big|\big(H \,- \stackrel{\,\leftarrow}{H}\!\big)\big| \tilde{\psi} (t) \big>$ \footnote{The respective anticausal continuity equations are obtained by Hermitian conjugation.}. The right hand side of these equations leads for standard --- even non-Hermitian --- Hamilton operators being quadratic in the momentum at most to (spacial) surface terms!
The corresponding non-vanishing --- in general {\em complex} --- densitities $\big<\!\big< \tilde{\psi} (t) \big| \psi (t) \big>$ and  $\big<\!\big< \psi (t) \big| \tilde{\psi} (t) \big>$ being in the standard way related to --- in general {\em complex} --- conserved charges may be interpreted as {\em complex} probability densities \cite{berggren1970} which replace Born's \cite{Pais:1982we} suggested ansatz $\big< \psi (t) \big| \psi (t) \big>$ for a {\em real} probability density.

\subsection{The (Anti)Causal Schr\"odinger Theory in Holomorphic Representation}
As in the traditional formulation of QM it is now natural to look for a spacial representation of the ``representation free'' Harmonic Oscillator introduced above. The particular complication induced by the existence of {\em two} types of annihilation operators~\mbox{($a$, $c$)} and creation operators ($c^+$, $a^+$) indicating a {\em doubling} of the degrees of freedom compared to the traditional Harmonic Oscillator is overcome by replacing the originally {\em real} spacial variable $x$ by a {\em complex} variable $z$ and its complex conjugate $z^\ast$ \footnote{M.\ Znojil \cite{Znojil:2001} achieves a complexification of a real coordinate $x$ by an overall shift $x\rightarrow x-i\delta$.}. This replacement of functions of one real argument $f(x)$ ($x\in{\rm I\!R}$) by respective functions $f(z,z^\ast)$ of complex arguments $z,z^\ast\in{\rm\bf C}$ is well known and used in complex analysis \cite{ablowitz2003} under the terminology ``holomorphic representation'' (e.g.\ Ref.\ \cite{Faddeev:1980be}). Replacing the right and left eigenstates $\big|x\big>$ and $\big<x\big|$ of the position operator by respective states in holomorphic representation $\big|z,z^\ast\big>$ and $\big<\!\big<z,z^\ast\big|$, we can --- analogously to traditional QM --- denote e.g.\ the four causal Schr\"odinger equations in their holomorphic representation by:
\begin{eqnarray} + i \,\partial_t \,\big<\!\big<z,z^\ast\big|\psi(t)\big> & = & \int dz^\prime dz^{\prime\ast} \, \big<\!\big<z,z^\ast\big|H\big|z^\prime,z^{\prime\ast}\big>\big<\!\big<z^\prime,z^{\prime\ast}\big|\psi(t)\big> \; , \nonumber \\
- i \,\partial_t \,\big<\!\big<z,z^\ast\big|\tilde{\psi}(t)\big> & = & \int dz^\prime dz^{\prime\ast} \, \big<\!\big<z,z^\ast\big|H\big|z^\prime,z^{\prime\ast}\big>\big<\!\big<z^\prime,z^{\prime\ast}\big|\tilde{\psi}(t)\big> \; , \nonumber \\
- i \,\partial_t \,\big<\!\big<\tilde{\psi}(t)\big|z,z^\ast\big> & = & \int dz^\prime dz^{\prime\ast} \, \big<\!\big<\tilde{\psi}(t)\big|z^\prime,z^{\prime\ast}\big>\big<\!\big<z^\prime,z^{\prime\ast}\big|\!\stackrel{\,\leftarrow}{H}\big|z,z^\ast\big> \; , \nonumber \\
+ i \,\partial_t \,\big<\!\big<\psi(t)\big|z,z^\ast\big> & = & \int dz^\prime dz^{\prime\ast} \, \big<\!\big<\psi(t)\big|z^\prime,z^{\prime\ast}\big>\big<\!\big<z^\prime,z^{\prime\ast}\big|\!\stackrel{\,\leftarrow}{H}\big|z,z^\ast\big> \; .
\end{eqnarray}
The spacial integration contours $\int dz^\prime\, dz^{\prime\ast}$ are to be performed such that there holds the generalized completeness relation $\int dz \, dz^\ast \,\big|z,z^\ast\big>\big<\!\big<z,z^\ast\big| = \mbox{\bf 1}$. 
Inversely, by the used notation it is understood that there holds a generalized orthogonality relation $\big<\!\big<z,z^\ast\big|z^\prime,z^{\prime\ast}\big> = \delta(z-z^\prime)\,\delta(z^\ast - z^{\prime\ast})$, in which the $\delta$-distributions for complex arguments are assumed to exist with respect to the chosen integration contours mentioned above. The holomorphic representation of the (Anti)Causal {\em Bosonic} Harmonic Oscillator of Section \ref{acho1} is provided by a translational invariant Hamilton operator $\big<\!\big<z,z^\ast\big|H\big|z^\prime,z^{\prime\ast}\big>= H(z,z^\ast) \,\big<\!\big<z,z^\ast\big|z^\prime,z^{\prime\ast}\big>$ given by $H(z,z^\ast) = -\, \frac{1}{2\, M} \,\frac{d^2}{dz^2} + \frac{1}{2} \; M \, \omega^2 \, z^2 -\, \frac{1}{2\, M^\ast} \, \frac{d^2}{dz^{\ast 2}} + \frac{1}{2} \; M^\ast \, \omega^{\ast 2} \, z^{\ast 2}$. 
Obviously there holds the following correspondence between annihilation/creation operators in the representation free case and in the holomorphic representation (with $p = - i \, d/dz$, $p^\ast = - i \, d/dz^\ast$): $c^+ \leftrightarrow \left( p + i\, M\,\omega \, z \right)/\sqrt{2\,M\,\omega} $, $a \leftrightarrow \left( p - i\, M\,\omega \, z \right)/\sqrt{2\,M\,\omega}$, $c \leftrightarrow \left( p^\ast -  i\, M^\ast\,\omega^\ast \, z^\ast \right)/\sqrt{2\,M^\ast\,\omega^\ast}$, $a^+ \leftrightarrow \left(p^\ast + i\, M^\ast\,\omega^\ast \,z^\ast \right)/\sqrt{2\,M^\ast\,\omega^\ast}$.

By this correspondence it is possible to construct the {\em normal} eigensolutions of the stationary Schr\"odinger equation $H(z,z^\ast )\,\,\big<\!\big< z,z^\ast\big|n,m\big> = E_{n,m} \,\big<\!\big< z,z^\ast\big|n,m\big>$ as
$\big<\!\big< z,z^\ast\big| n,m\big> =
  i^{\,n+m} \sqrt{|M\omega|/(2^{\,n+m} \, n!\,  m! \, \pi)} \, \exp\left(-\,\frac{1}{2} ( \xi^2 + \xi^{\ast\,2})\right) \, H_n(\xi )\,H_m(\xi^\ast)$ with $\xi = z \; \sqrt{M \,\omega}$. The inverse oscillator length $\sqrt{M \,\omega}$ is here {\em complex} valued. $H_n(\xi)$ and $H_m(\xi^\ast)$ are Hermite polynomials with complex arguments \footnote{Note that (anti)causal orthonormality relations $\int d\xi \, \exp(-\xi^2) \,H_n (\xi) \,H_{m} (\xi) = 2^n n! \sqrt{\pi}\,\delta_{n m}$ and $\int d\xi^\ast \, \exp(-\xi^{\ast 2}) \,H_n (\xi^\ast) \,H_{m} (\xi^\ast) = 2^n n! \sqrt{\pi}\,\delta_{n m}$
are quite different from the ones of Glauber coherent states \cite{Glauber:1963tx,Faddeev:1980be}: $\int d\xi \,d\xi^\ast \exp(-|\xi|^2) \,\psi^\ast_n (\xi) \, \psi_m(\xi^\ast)/(2\pi i) = \delta_{n m}$ with $\psi_n(\xi)=\xi^n/\sqrt{n!}$.}\footnote{Within the holomorphic representation it is easy to understand that there exist two distinct time-reversal operations ${\cal T}$ and $T$ discussed in Ref.\ \cite{Kleefeld:2004jb}, the former being just the anti-unitary Hermitian conjugation, the latter interchanging initial and final states.}.
\subsection{Reconciling Causality, Poincar\'{e} Invariance, Analyticity, Locality in NHQT}
While conceptions like causality and Poincar\'{e} invariance are fundamental principles of physics, which should not be given up without need in theories describing nature, the features of analyticity and locality seem to be required merely for calculatory convenience. Fortunately we observe that a restoration of causality within a Poincar\'{e} invariant QT goes hand in hand with a restoration of analyticity and locality.
One mandatory argument discussed in Ref.\ \cite{Kleefeld:2004jb} to extend Hermitian QT to NHQT has been the incompatibility of causality with the traditional anti-particle concept. We will illustrate here only the point that all fields (or states) --- including the ones serving as asymptotic states --- in an interacting theory should be considered as fields with {\em complex-valued mass} (i.e.\ like (anti)Gamow states) and should follow a {\em postulate} specified below to allow simultaneously causality, Poincar\'{e} invariance, analyticity, and locality. Let's decompose e.g.\ the Lorentz invariant (non-Hermitian) causal and anticausal KG fields into their Hermitian (``shadow'' \cite{Kleefeld:2004jb}) components $\phi_{(1)}(x)$ and $\phi_{(2)}(x)$, i.e.\ $\phi(x) = ( \phi_{(1)}(x) + i \, \phi_{(2)} (x))/\sqrt{2}$, $\phi^+(x) = ( \phi_{(1)}(x) - i \, \phi_{(2)} (x))/\sqrt{2}$. Then the first Lagrangean in Eq.\ (\ref{kglag1}) can be rewritten as ${\cal L} (x) =  
\frac{1}{2} \,\big( (\partial \phi_{(1)} (x) )^2  - \mbox{Re}[M^2]  
(\phi_{(1)} (x))^2 \big) - \frac{1}{2} \,\big( (\partial \phi_{(2)} (x) )^2 - \mbox{Re}[M^2]  (\phi_{(2)} (x))^2 \big) + \mbox{Im}[M^2]  \; \phi_{(1)} (x) \,\phi_{(2)} (x)$.
We observe that one shadow field has {\em positive} norm, the other has {\em negative} norm displaying the underlying {\em indefinite metric}. Secondly we recall that shadow fields are not described by causal or anticausal propagators, but by {\em acausal} linear combinations which reduce for quasi-real masses to {\em principal value propagators} or {\em $\delta$-distributions}. Finally we see that the Lagrangean is not diagonal in the shadow fields, while the interaction term is proportional to the imaginary part of $M^2$. If one would remove the interaction term, one would introduce interactions between causal and anticausal fields (e.g. $\phi(x) \, \phi^+(x)$ in the KG theory, or $z z^\ast$ in the holomorphic representation of the Bosonic Harmonic Oscillator) leading not only to a loss of analyticity, yet also to a loss of causality due to interactions between causal and anticausal states. To obtain an acceptable S-matrix for his Complex-Ghost Relativistic Field Theory {\em containing interactions} between causal and anticausal KG fields N.\ Nakanishi was forced to introduce {\em Lorentz non-invariant} functions to regulate integrals. Hence he observed for this case also a loss of {\em Lorentz invariance} \cite{Nakanishi:1971jj,Kleefeld:Nakanishi:wx} (See also \cite{Lee:1971ix,Kleefeld:Nakanishi:1972pt,Gleeson:1972xj}!). Yet there is a simple way to cure the lack of Lorentz invariance, analyticity, causality and locality: it is our POSTULATE (also for e.g.\ (anti)causal Dirac theory) that there should be {\em no interaction terms between causal fields (e.g.\ $\phi(x)$) and anticausal fields (e.g.\ $\phi^+(x)$) in the Lagrange density} \footnote{Even asymptotic states have to be treated like complex ghosts with complex-valued mass containing an infinitesimal imaginary part, which is quite consistent with all the features of in- and out-states in a non-stationary description of standard Hermitian QT. Note that asymptotic states of strictly real mass (or energy) would be a superposition of causal and anticausal states, which would typically couple to intermediate states such that they violate our postulate!}!

\section{Applications}
\subsection{The ``Shifted'' (Anti)Causal Harmonic Oscillator}
In 1997 C.M.\ Bender \& S.\ Boettcher \cite{Bender:1998ke} (See also M.\ Znojil \cite{Znojil:1999qt,Znojil:2001}!) used the non-Hermitian Hamilton operator $H=p^2 + x^2 + i\,x= p^2 + (x+i/2)^2 +1/4$ obtained from a Harmonic Oscillator shifted to a complex space point $x=-i/2$ as an example to show that its spectrum $E_n=(2\,n+ 1) + 1/4 = 2\,n + 5/4$ can be indeed real due to the underlying ``PT-symmetry''. We shall study here this shift in the holomorphic representation, in which the unshifted oscillator is described by the Hamilton operator $H(z,z^\ast) = -\, \frac{p^2}{2\, M} + \frac{1}{2} \; M \, \omega^2 \, z^2 -\, \frac{p^{\ast\,2}}{2\, M^\ast} + \frac{1}{2} \; M^\ast \, \omega^{\ast 2} \, z^{\ast 2}$. Without changing its spectrum the Hamilton operator is shifted from $(z,z^\ast)$ to \mbox{$(z+\alpha,z^\ast+\beta^\ast)$} by a standard equivalence transform $H(z+\alpha,z^\ast+\beta^\ast) = U_{z,z^\ast}(\alpha,\beta^\ast) \; H(z,z^\ast) \; U^{-1}_{z,z^\ast}(\alpha,\beta^\ast)$ with $U_{z,z^\ast}(\alpha,\beta^\ast)=\exp\big( i \, ( \alpha \, p + \beta^\ast \, p^\ast ) \big)$ and $U^{-1}_{z,z^\ast}(\alpha,\beta^\ast)\equiv U_{z,z^\ast}(-\alpha,-\beta^\ast)$. The ``shifted'' Hamilton operator $H(z+\alpha,z^\ast+\beta^\ast) = H(z,z^\ast) + M \, \omega^2 \, \alpha \left(z+\frac{\alpha}{2}\right)  \; +  M^\ast \, \omega^{\ast \,2} \, \beta^\ast \big(z^\ast + \frac{\beta^\ast}{2}\big)$ will be $P{\cal T}$-symmetric (i.e.\ $H(z+\alpha,z^\ast+\beta^\ast)^{P{\cal T}}=H(z+\alpha,z^\ast+\beta^\ast)$) for $\alpha=-\beta \Leftrightarrow \alpha^\ast=-\beta^\ast$. It will be Hermitian/${\cal T}$-symmetric (i.e.\ $H(z+\alpha,z^\ast+\beta^\ast)^{{\cal T}}=H(z+\alpha,z^\ast+\beta^\ast)$) for $\alpha=\beta \Leftrightarrow \alpha^\ast=\beta^\ast$. Bender's example ``$H=p^2 + x^2 + i\,x\,$'' is essentially obtained by $H(z+i\,\gamma,z^\ast+i\,\gamma^\ast)$ with $\gamma = +1/2$. For $\omega\not=\omega^\ast$ --- even being strictly $P{\cal T}$-symmetric --- only some part of its spectrum $E_{n,m}=\omega \,(n+\frac{1}{2})+\omega^\ast (m+\frac{1}{2})$ with $n,m\in{\rm I\!N}$${}_0$ is real, namely if $n=m$.
\subsection{Non-Hermitian Supersymmetry}
Non-Hermitian supersymmetric Hamilton operators with PT-symmetry have been investigated already for quite some time (See e.g.\ Ref.\ \cite{Znojil:2000nh}!). We want here to go one step further and
consider the non-PT-symmetric Hamilton operator of an (Anti)Causal Supersymmetric Harmonic Oscillator, being the sum of a causal Bosonic and
Fermionic Harmonic Oscillator with equal {\em complex} frequency $\omega_C$, and an anticausal Bosonic and
Fermionic Harmonic Oscillator with equal {\em complex} frequency $\omega_A$, i.e. $H = 
\frac{1}{2} \; \omega_C \, \{c^+, a\, \}
+\frac{1}{2} \; \omega_C \, [\,d^+, b\, ]
+ \frac{1}{2} \; \omega_A^\ast \, \{a^+, c\, \}
+ \frac{1}{2} \; \omega_A^\ast \, [\, b^+, d\, ] =
\omega_C \; (c^+ a+ d^+ b )
+\omega_A^\ast \, (a^+ c+ b^+ d )=\omega_C \; \{ Q_+ , Q_-\}
+\omega_A^\ast \, \{ Q^+_- , Q^+_+\}$. Here we introduced supercharges $Q_+ = a \,d^+$, $Q_+^+ = d \,a^+$, $Q_- = c^+ b$, $Q_-^+ = b^+ c$, which are as usual nilpotent ($Q_\pm^2 = ( Q_\pm^+)^2 = 0$), but {\em not} related by Hermitian conjugation, as $(Q_\pm)^+ \not= Q_\mp$. As usual in supersymmetric systems the positive and negative contributions to the Bosonic and Fermionic vacuum energy cancel.
The Hamilton operator is easily diagonalized by the {\em normal} right eigenstates $\big| n_B,n_F;\bar{n}_B,\bar{n}_F\big>=(c^+)^{n_B} (d^+)^{n_F} (a^+)^{\bar{n}_B} (b^+)^{\bar{n}_F}\big|0\big>/\sqrt{n_B!\,\bar{n}_B!}$ or left eigenstates $\big<\!\big< n_B,n_F;\bar{n}_B,\bar{n}_F\big|=\big<\!\big< 0 \big| d^{\,\bar{n}_F} c^{\bar{n}_B} b^{n_F} a^{n_B}/\sqrt{n_B!\,\bar{n}_B!}$ yielding the (complex) eigenenergies $E_{n_B,n_F;\bar{n}_B,\bar{n}_F} = \omega_C (n_B+n_F) + \omega_A^\ast (\bar{n}_B+\bar{n}_F)$ (with $n_B$, $\bar{n}_B \in {\rm I\!N}$${}_0$ and  $n_F$, $\bar{n}_F \in \{0,1\}$) \footnote{Application of supercharges interrelates as usual distinct eigenstates of the Hamilton operator belonging to equal eigenvalues. 
$P{\cal T}$-symmetry is restored by setting $\omega\equiv\omega_C=\omega_A$.}. 
\subsection{Towards a Theory of Strong Interactions without Gluons}
Mysteriously, the Quark-Level Linear Sigma Model \cite{Delbourgo:1993dk} (QLL$\sigma$M) has been a rather successful theory to describe various experimental facts involving hadronic physics at low and intermediate energies (See e.g.\ \cite{Kleefeld:2001ds} and references therein!). Guided by this observation we ``mapped'' \cite{Kleefeld:2002au} in 2002 by a simplistic argument the Lagrangean of QCD into a Lagrangean of a QLL$\sigma$M which is supposed to describe quark-quark scattering equally well {\em at high energies}.
The unexpected result has been a {\em non-Hermitian} QLL$\sigma$M Lagrangean, which is {\em asymptotically free} due to a purely {\em imaginary} Yukawa coupling $g$ between scalar mesons and quarks, the PT-symmetry of which suggests in correspondence to C.M.\ Bender's ``physical'' $i\,\phi^3$-theory \cite{Bender:1998ke,Bender:2004by} a {\em real spectrum} of the respective Hamilton operator \footnote{By exact cancellation of quadratic divergencies in selfenergies of (pseudo)scalar mesons one can establish a relation between the quartic coupling $\lambda$ of the QLL$\sigma$M and the Yukawa coupling g of the form $g^2 \propto  \lambda +O(\lambda^2)$. To one loop (i.e. by skipping $O(\lambda^2)$) one can deduce that a {\em imaginary} Yukawa coupling $g$ should be accompanied by a {\em negative} quartic coupling $\lambda$. This explains partially the intimate relation between C.M.\ Bender's \cite{Bender:1998ke,Bender:1999ek} ``physical'' $i\,\phi^3$-theory and K.\ Symanzik's \& C.M.\ Bender's \cite{Bender:1999ek,Bender:1998ke} ``physical'' $-\phi^4$-theory. Moreover suggests the (non-perturbative) two-loop relation between $g$ and $\lambda$ that at least one of these couplings has to be complex-valued for $\lambda$ finite.}! 

This work has been supported by the
{\em Funda\c{c}\~{a}o para a Ci\^{e}ncia e a Tecnologia} \/(FCT) of the {\em Minist\'{e}rio da Ci\^{e}ncia e da Tecnologia (e do Ensinio Superior)} \/of Portugal, under Grants no.\ PRAXIS
XXI/BPD/20186/99, SFRH/BDP/9480/2002, and POCTI/\-FNU/\-49555/\-2002.

\end {document}